\documentclass[aps, prd, twocolumn, tightenlines, notitlepage, superscriptaddress, nofootinbib, preprintnumbers, floatfix, showkeys,10pt]{revtex4-2}

\usepackage{graphicx}
\usepackage{xcolor}
\usepackage[utf8]{inputenc}
\usepackage{amsmath}
\usepackage{tabularx}
\def\beqn#1{\begin{equation}\label{#1}}
\def\eeqn{\end{equation}}
\def\beqa#1{\begin{eqnarray}\label{#1}}
\def\eeqa{\end{eqnarray}}

\def\be{\begin{equation}}

\def\ee{\end{equation}}
\def\barr{\begin{array}}
\def\earr{\end{array}}

\def\nn8{\nonumber\\[2pt]}

\def\ed{\end{document}}

\begin{document}
 
 \title{Coherent and incoherent antineutrino scattering on stable even-even isotopes of molybdenum detectors}


\author{T.S. Kosmas}
\email{Corresponding author, {\it E-mail address:} hkosmas@uoi.gr}
\affiliation{Division of Theoretical Physics, University of  Ioannina, GR 45110 Ioannina, Greece,}
\author{R. Sahu}
\email{{\it E-mail address:} rankasahu@gmail.com} 
\affiliation{NIST University, Institute Park, Berhampur 761 008, Odisha, India,}
\author{V.K.B. Kota}
\email{{\it E-mail address:} vkbkota@prl.res.in}
\affiliation{Physical Research Laboratory, Ahmedabad 380 009, India.}

\begin{abstract}
The recent observations of the coherent neutrino- and antineutrino-nucleus scattering have 
opened up a plethora of opportunities to probe physics within standard and non-standard theories 
of the electroweak interactions.
In the present article, our goal is to explore the possibility of using the molybdenum material 
as detection medium for coherent and incoherent antineutrino- and neutrino-Mo scattering in the 
ongoing and future coherent elastic neutrino-nucleus scattering
(CE$\nu$NS) experiments by using relevant (anti-)neutrino beams as e.g. 
stopped pion-decay neutrino beams, reactor antineutrino beams, astrophysical (solar or supernova) 
(anti)neutrino beams, etc. Our present coherent and incoherent scattering cross sections of 
Mo isotopes with neutrinos and antineutrinos are based on the deformed shell model (DSM) that 
has been previously employed for studying similar processes. On the other hand, in the past, 
CE$\nu$NS events obtained with this model provided us with better fits to experimental data of 
COHERENT experimental data compared to phenomenological form factors.
\end{abstract}

\maketitle

\section{Introduction}

The coherent elastic neutrino-nucleus scattering (CE$\nu$NS) is a Standard Model rare event 
(due to the very weak neutrino-matter interaction) process, governed by the electroweak 
interaction, that was predicted long ago (1974) by Freedman \cite{Freedman-1974}. In recent
neutrino ($\nu$) and antineutrino ($\tilde{\nu}$) detection experiments, due to the very
weak neutrino-nucleus interaction, a tonne or multi-tonne mass scale of the detection
medium is required. In the case of the CE$\nu$NS experiments \cite{COHERENT-Science} 
extremely low-energy thresholds 
(of sub-keV scale) nuclear detectors are employed which drastically enhance the event 
rates. CE$\nu$NS has proven to be a difficult process to detect due to the fact that the
deposited recoil-energy in the nuclear detector is small (of the order of keV). 

The first detection of CE$\nu$NS was announced by the COHERENT collaboration
\cite{COHERENT-PRL,COHERENT-Ar} in 2017, i.e. forty three years after its first prediction 
by Freedman \cite{Freedman-1974}. The first COHERENT experiment used sodium doped
CsI detectors with a decay at rest (DAR) pion neutrino-source while the second
one used Ar target. 

The COHERENT measurement of CE$\nu$NS events motivated a steady increase in the
experimental sensitivity through the XENONnT \cite{XENON}, LZ \cite{LZ-23}, and
PandaX-4T \cite{PandaX-4T}
collaborations which reported their first observations of nuclear recoils from
solar $^{8}$B neutrino source via CE$\nu$NS on Xenon and Germanium detectors. 
By assuming the non existence of new physics,
these results provide a measurement of the component $^{8}$B of solar neutrino flux 
and motivated investigations aiming at probing new physics 
via non-standard neutrino interactions (NSI), light mediators,as well as the 
determination of the weak mixing angle at low momentum transfer
\cite{Kosmas-Valle-2015,Kosmas-Valle-2017,Papoulias-tsk-2018,Aristizabal,Li-Song}.

Recently, the CONUS experiment \cite{CONUS-PRL-2023,CONUS} measured CE$\nu$NS 
events by using a high-purity Ge detector 
with a total active mass $(3.73\pm 0.02)$ kg, positioned at a distance 17.1 m from
the core center of the nuclear power plant in Brokdorf, Germany, which provided
a reactor antineutrino flux (up to $2.3\times 10^{13} cm^{-2} s^{-1}$). The
ionization energy threshold achieved in this Ge detector was very low (0.21 keV),
while the combined likelihood fit performed for this experiment provided an upper
limit of 143 antineutrino events (at 90\% confidence level). According to CONUS
collaboration, this number of events is less than a factor of 2 larger than
the predicted events by the SM \cite{CONUS-PRL-2023}.

In the case of the CONUS+ experiment \cite{CONUS+,C+new}, high purity Germanium 
crystals (with 160-180 eV threshold) were employed to measure reactor 
antineutrino-Ge scattering events for the first time. The reactor antineutrinos 
are of low-energy (up to about 8-10 MeV). The number of events measured by the 
CONUS+ experiment was $(395 \pm 106)$ (during operational time of 119 days 
at the nuclear power plant in Leibstadt, Switzerland) with a statistical 
significance 3.7 $\sigma$ \cite{C+new}. The predicted number of events by the
Standard Model of the electroweak interactions is equal to $(347\pm 59)$
which means that there is room for new physics beyond the Standard Model
\cite{Papoulias-tsk-2018,conusp25,conu-appl,solarnu}.

The possible detection of CE$\nu$NS using both terrestrial and astrophysical 
neutrino-sources has inspired new constraints on the physics beyond the Standard 
Model (BSM) and other significant implications for particle physics, astrophysics, 
and nuclear physics; see for example \cite{Popu-new,bsm-CMS,Giunti-PLB-2025}. 

Nowadays, in the intense experimental effort ongoing today around the world, 
the CE$\nu$NS process is studied with a variety of neutrino sources and detector 
technologies by employing various detection materials 
~\cite{CEvNS-Anti-nu,CONNIE,v-GeN,MINER,Chooz,v-cleus,TEXONO,RICOCHET}
(research from a theoretical viewpoint is also increasing 
\cite{JPG2024,MDP1,MDP2,cRPA,Tsakstara-tsk,qrpa-1,mpqm-1,mqrpa-2,SM-PHS,VegBon,cenu-sm,Verg-2,nth-1,nth-2,nth-3,nth-4,nth-5,nth-6}). 
For example, new experiments have been designed, like the NUCLEUS experiment at 
TU Munich, Germany, aiming for the measurement of CE$\nu$NS in a nuclear reactor
using reactor neutrinos (there, the neutrino fluxes have high intensity) at the, 
relatively, very low energy range (up to 10 MeV) of the nuclear power 
plant in Chooz, France. 

The NUCLEUS experiment is going to utilize dedicated cryogenic detectors with nuclear 
recoil energy thresholds estimated to be around 20 eV. This is the lowest used in the 
topic of CE$\nu$NS. At present, the NUCLEUS setup is in the stage of construction (in the 
shallow underground laboratory at TUM) and commissioning, and it is expected that the 
experiment will be moved to France next year (a technical run is anticipated to take place
in 2026). The scientific potential with several technological details of this experiment
can be found, e.g., in \cite{v-cleus,NUCLEUS-Coll-2023}.
 
Our aim in this paper is to share this effort by providing useful theoretical predictions 
for both the coherent and incoherent scattering of neutrinos and antineutrinos on the 
even-even molybdenum isotopes. Toward this 
aim, we utilize the well known Donnelly-Walecka multipole decomposition method based on 
the deformed shell model. The latter method, has been adopted previously to study
rare event processes like the muon-to-electron conversion \cite{Kosmas-2003}, direct 
dark matter detection \cite{AHEP_2018} as well as detection of CE$\nu$NS in several 
nuclear isotopes \cite{PLB_2020}. 

The Molybdenum (with atomic number Z=42) is considered a promising detection medium
for rare event processes as is the CE$\nu$NS and has been recently employed by
extremely sensitive experiments like the NEMO neutrinoless double beta decay
\cite{Arnold-NEMO-2015} and the MOON direct dark matter detection \cite{Ejiri-2007} 
experiments.  As it is known \cite{Mo-abun}, molybdenum has seven stable isotopes 
with N=50, 52, 53, 54, 55, 56, 58, and abundances as follows: $^{92}$Mo (15.86\%),
$^{94}$Mo (9.12\%), $^{95}$Mo (15.70\%), $^{96}$Mo (16.50\%), $^{97}$Mo (9.45\%), 
$^{98}$Mo (23.75\%), and $^{100}$Mo (9.62\%). 

In our present work, we have considered 
all the even-even isotopes $^{92,94,96,98,100}$Mo.
Recently, the AMoRE collaboration has been searching for neutrinoless double beta decay 
of the $^{100}$Mo using molybdate scintillating crystals (100 kg of enriched
$^{100}$Mo) via low temperature thermal calorimetric detection 
\cite{AMoRE-PRL,AMoRE-Eur-Phys-C}. 

We note that CE$\nu$NS detectors could also be used as dark matter (DM) detectors as
has been pointed out in Ref. \cite{Goodman-Witten} where the anticipation of similar
experimental challenges is discussed.

Within the context of the multipole expansion Donnelly-Walecka method 
\cite{Donnelly-1, Donnelly-2}, 
we calculate coherent elastic (anti)neutrino-Mo scattering cross sections, for
ground state to ground state ($\vert gs\rangle\to \vert gs\rangle$) transitions
as well as incoherent (inelastic) cross sections where the nucleus from the ground 
state transitions to an excited (higher energy)
state ($\vert gs\rangle\to \vert f\rangle$), where $\vert f\rangle$ represents
a low-lying (up to about 10 MeV) excited state. By using these cross sections total 
cross sections (total = coherent + incoherent) are readily obtained. Moreover,
the important ratio of the coherent over the total cross section,
$\rho = coherent/total$, for each Mo isotope is calculated. The latter quantity 
provides the portion of the total rate (unknown and appreciably difficult to measure) 
represented by the measured coherent rate in the CE$\nu$NS experiments. 

The required ground 
($\vert gs\rangle$) and final ($\vert f\rangle$) states for the even-even 
Mo isotopes, $^{92,94,96,98,100}$Mo, are deduced through nuclear structure
calculations by utilizing the deformed shell model (DSM) as is explained below. 
It should be noted that, the DSM eigenfunctions, with the corresponding eigenenergies 
for the ground and various
excited states in $^{94,96,98,100}$Mo have been determined in our previous
publications \cite{JPG2024,MDP1,MDP2}. Here, we will first present the DSM results 
for the remaining $^{92}$Mo isotope and, then, we will proceed with a comparative 
analysis of these results through the computed ratio $\rho = \text{coherent/total}$
for all even-even Mo isotopes. We note that the results for neutrino scattering 
off nuclei differ from those of antineutrino scattering due to the cross-term
"polar vector times axial vector" that appears in the interaction Hamiltonian
\cite{Tsakstara-tsk}.

In the rest of the paper the material is organized as follows. At first (Sect.
2), the neutrino-nucleus scattering cross section formalism based on the
Donnelly-Walecka method is briefly summarized. Also the deformed shell model
(DSM) relying on Hartree-Fock deformed intrinsic states with angular momentum 
projection and band mixing, which provides the low-lying nuclear states required
for our present calculations is briefly described. Following this, in Sect. 3, 
initially, cross section results for $^{92}$Mo are presented in detail. Then, 
in Sect. 4, a comparative analysis of the coherent
and incoherent scattering cross section results of neutrinos and antineutrinos
on the even-even molybdenum isotopes is presented and discussed in conjunction
with the important quantity $\rho$. Finally, in Sect. 5, the conclusions 
extracted from the present study are briefly summarized. 
 
\section{Formalism of neutral-current $\nu$-nucleus reactions}
\subsection{Neutrino-nucleus scattering cross section within Donnelly-Walecka method}

The relevant formulation that provides the neutral current $\nu$-nucleus scattering
differential cross sections as a function of the incoming neutrino energy 
$\varepsilon_{\nu}$ have been discussed in detail earlier in Refs. 
\cite{Tsakstara-1,Donnelly-1,Donnelly-2,JPG2024}. A few important steps are 
given below for the sake of completeness.
In the low energy domain with the leptonic and hadronic currents represented
by $\hat{j}^{lept}_\mu$ and $\hat{\mathcal{J}}^\mu$, respectively, 
the Standard Model weak interaction neutrino-nucleus Hamiltonian $H_I$ 
is written in the effective current-current form as
\begin{equation}
\hat{H}_I=-\frac{G}{\sqrt{2}}\int d^3 \mathbf{x} \ \hat{j}^{lept}_\mu(\mathbf{x})
        \hat{\mathcal{J}}^\mu (\mathbf{x}) \, .
        \label{eq.1}
\end{equation}
In the above definition, $G$ is the Fermi weak coupling constant.
The double differential cross section for the scattering of low energy 
neutrinos from even-even Molybdenum isotopes for a transition from the 
initial state $\mid i\rangle$ represented by $\mid J_i, M_i\rangle$ to the final 
state $\mid f\rangle$ represented by $\mid J_f, M_f\rangle$ can be written as
\begin{equation}
\frac{d^2\sigma_{i\rightarrow f}}{d\Omega dw} = (2\pi)^4\varepsilon^2_f
        \sum_{s_f, s_i, M_f M_i} \displaystyle{\frac
        {       \mid\langle f \mid \hat{H}_I
        \mid i\rangle \mid^2}{(2J_i+1)}  }
        \label{eq.10}   
\end{equation}
where the sums are over the initial and final spin states ($s_i, s_f$) and magnetic 
quantum numbers ($M_i, M_f$), respectively. Applying a multipole analysis on the 
weak hadronic current following
Donnelly-Walecka method \cite{Donnelly-1,Donnelly-2}, the neutrino-nucleus scattering 
cross section for the excitation energy $\omega$ of the target nucleus becomes
\begin{widetext}
\begin{equation}
\frac{d^2\sigma_{i\rightarrow f}}{d\Omega d\omega}(\phi,\theta,\omega ,\varepsilon_i) =
        \delta(E_f-E_i-\omega) \frac{2G^2\varepsilon^2_f \cos^2(\theta/2)}
        {\pi (2J_i+1)}\left[C_V+C_A\mp C_{VA}\right]
        \label{eq.11}
\end{equation}
\end{widetext}
Because of the energy conservation, $\omega$ is equal to the difference of
the initial and final states of the nucleus. It is also equal to the difference
of the incoming and outgoing energies $\varepsilon_i$ and $\varepsilon_f$ of the neutrinos,
i.e. $\omega=E_f - E_i=\varepsilon_i - \varepsilon_f$.
In Eq. \ref{eq.11}, the $(-)$ sign corresponds to the scattering of the neutrinos and the
$(+)$ to the scattering of the antineutrinos.

The terms $C_V$ and $C_A$ in Eq. \ref{eq.11} include a summation over the contributions 
coming from the polar-vector and axial-vector multipole operators, respectively and are 
given by
\begin{widetext}
\begin{equation}
        \begin{array}{lll}
                C_{V(A)} & = &\displaystyle{\sum^\infty_{J=0}} \mid \langle J_f\mid\mid \hat{M}^{(5)}_J(q)
+\displaystyle{\frac{w}{q}} \hat{L}^{(5)}_J(q)\mid\mid J_i\rangle\mid^2\\
                &&+\displaystyle{\sum^\infty_{J=1}}\left(\displaystyle{-\frac{q^2_\mu}{2q^2}+\tan^2\frac{\theta
                }{2}}\right)\left[\mid\langle J_f\mid\mid|\hat{T}^{mag(5)}_J(q)
                \mid\mid J_i\rangle\mid^2 \right.
\left.+\mid\langle J_f\mid\mid\hat{T}^{el(5)}_J(q)\mid\mid
                J_i\rangle\mid^2\right]
        \end{array}
        \label{eq.13}
\end{equation}
\end{widetext} 
Similarly, the interference term $C_{VA}$ in Eq. \ref{eq.11} contains the product of transverse polar-vector 
and transverse axial-vector matrix elements and is given by
 
\begin{equation}
        \begin{array}{lll}
C_{VA}&=&2\tan \displaystyle{\frac{\theta}{2}}\left(\displaystyle
 {-\frac{q^2_\mu}{q^2}+\tan^2\frac{\theta }{2}}\right)^{1/2}\\
&&\times\displaystyle{\sum^\infty_{J=1}}Re\langle J_i\mid\mid\hat{\mathcal{T}}^{mag}_J(q)\mid\mid J_i\rangle
\langle J_f\mid\mid\hat{\mathcal{T}} ^{el}_J(q)\mid\mid J_i\rangle^*
        \end{array}
        \label{eq.14}
\end{equation}

In the above equations, the superscript 5 refers to the axial vector components of the hadronic current and 
they are given in the Appendix. $C_{VA}$ contains contributions of $\hat{T}^{el}_J$ and $\hat{T}^{mag5}_J$
operators for normal parity transitions, while for abnormal parity ones it contains matrix elements of 
$\hat{T}^{mag}_J$ and $\hat{T}^{el5}_J$. 

The square of the four-momentum transfer and the three momentum transfer, the
scattering angle $\theta$ which occur in the above equations 
can be expressed as
\begin{equation}
        \begin{array}{l}
q_\mu^2\equiv q_\mu q^\mu = -4\varepsilon_i(\varepsilon_i-\omega)\sin^2(\theta/2) \\
q \equiv \mid\mathbf{q}\mid=\left[ \omega^2+4\varepsilon_i(\varepsilon_i-\omega)\sin^2(\theta/2)
                \right]^{1/2}
        \end{array}
\end{equation}
It is important to mention that in practice, we are interested in the differential cross section 
$d\sigma_{i \rightarrow f}/d\omega$ and this is obtained by integrating out $\phi$ and $\theta$ 
in Eq. (\ref{eq.11}). Note that the $C_v$, $C_A$ and $C_{VA}$ also depend on $\theta$.
The reduced matrix elements that enter in the formulas for $C_{V(A)}$ and $C_{VA}$ are evaluated 
within the deformed shell model (DSM) 
described ahead \cite{ks-book} for the nuclear structure part. Now we will briefly describe DSM.

\subsection{Deformed shell model}

The deformed shell model or DSM is based on Hartree-Fock (HF) deformed intrinsic states 
with angular momentum projection and band mixing. Over the years this model has been 
able to provide a good description of the properties of nuclei in the mass range 
$A=60-90$~\cite{ks-book} like the spectroscopy of $N=Z$ odd-odd nuclei with
isospin projection see for example~\cite{Srivastava-2017}, double beta decay
half-lives~\cite{Sahu-2013, Sahu-2015}, $\mu -e$ conversion in the 
field of the nucleus~\cite{Kosmas-2003} etc. In recent years, the model
has been applied to study neutrino-nucleus scattering, see for example our recent papers 
\cite{JPG2024,MDP1,MDP2}.
The details of this model have  been described in many of our earlier
publications, see for example~\cite{ks-book}. 

Assuming axial symmetry, for a given
nucleus, starting with a model space consisting of a given set of single
particle (sp) orbitals and effective two-body Hamiltonian (TBME + spe), the
lowest energy intrinsic states are obtained by solving the Hartree-Fock (HF)
single particle equation self-consistently. Excited
intrinsic configurations are obtained by making particle-hole excitations over
the lowest intrinsic state.  It is worth noting that intrinsic states $\chi_K(\eta)$ 
do not have definite angular momenta.  Hence states of good  angular momentum projected 
from an intrinsic state $\chi_K(\eta)$ can be  written in the form
\begin{equation}
\psi^J_{MK}(\eta) = \frac{2J+1}{8\pi^2\sqrt{N_{JK}}}\int 
d\Omega D^{J^*}_{MK}(\Omega)R(\Omega)| \chi_K(\eta) \rangle \, ,
\label{eqn.24}
\end{equation}
where $N_{JK}$ is the normalization constant given by
\begin{equation}
N_{JK} = \frac{2J+1}{2} \int^\pi_0 d\beta \sin \beta d^J_{KK}(\beta)
\langle \chi_K(\eta)|e^{-i\beta J_y}|\chi_K(\eta) \rangle  \, .
\label{eqn.25}
\end{equation}
In Eq.(\ref{eqn.24}), $\Omega$ represents the Euler angles ($\alpha$, $\beta$,
$\gamma$), while $R(\Omega)=\exp(-i \alpha  J_z) \exp(-i \beta J_y) \exp(-i \gamma J_z)$ 
represents the general rotation operator.  The good
angular momentum states projected from different intrinsic states are not in
general orthogonal to each other.  Hence they are orthonormalized and then 
band mixing calculations are performed.  The resulting eigenfunctions are of 
the form
\begin{equation}
\vert\Phi^J_M(\eta) \rangle \, =\,\sum_{K,\alpha} S^J_{K \eta}(\alpha)\vert 
\psi^J_{M K}(\alpha)\rangle \, .
\label{phijm}
\end{equation}
The reduced matrix elements occurring in Eqs. \ref{eq.13} and \ref{eq.14} are
evaluated using the band mixed wave functions $\Phi^J_M$ defined in Eq. \ref{phijm}. 
Now, we will present the DSM results for $^{92}$Mo,

\section{ Results and Discussion}

Initially, in our present calculations, we concentrated on the coherent and incoherent 
scattering cross sections (of neutrino and antineutrino) for the even-even molybdenum 
isotopes, focusing on the $^{92}$Mo which had not been studied in detail previously
\cite{JPG2024,MDP2}. Towards this purpose, we first derived and tested the DSM wave 
functions for the ground state and for the low-lying excited states (up to about 
$\omega =$ 15 MeV) of this isotope. Afterwards, by employing the wave functions deduced 
in our recent publications \cite{JPG2024,MDP2}, we computed individual cross sections of 
antineutrino scattering on the $^{94,96,98,100}$Mo isotopes, as described below. 

\subsection{ The nuclear structure of $^{92}$Mo isotope}

The effective nucleon-nucleon interaction employed for the $^{92}$Mo isotope
was the $GWBXG$ with the $^{66}$Ni as the 
inert core. The steps followed for its construction are described in detail in Ref. \cite{Dey}. 
This interaction has 
been successfully utilized in our previous studies \cite{JPG2024,MDP2} by assuming that the 
active orbits for the protons are the $0f_{5/2}$, $1p_{3/2}$, $1p_{1/2}$ and $0g_{9/2}$, with 
single-particle energies of $-5.322$, $-6.144$, $-3.941$ and $-1.250$ MeV, while for the neutrons 
the active orbits are $1p_{1/2}$, $0g_{9/2}$, $0g_{7/2}$, $1d_{5/2}$, $1d_{3/2}$, and $2s_{1/2}$. 
The single-particle energies for the first five neutron orbits are taken to be $-0.696$, $-2.597$, 
$5.159$, $1.830$, and $4.261$ MeV, respectively. The $2s_{1/2}$ neutron orbit produces low-lying 
large deformed solutions even though molybdenum isotopes are known to be weakly deformed. Hence, 
the effect of this orbit is eliminated by taking the corresponding neutron single-particle energy 
at a high value, just as in Refs. \cite{JPG2024,MDP2}.

\begin{figure}[ht]
\includegraphics[width=7.8 cm]{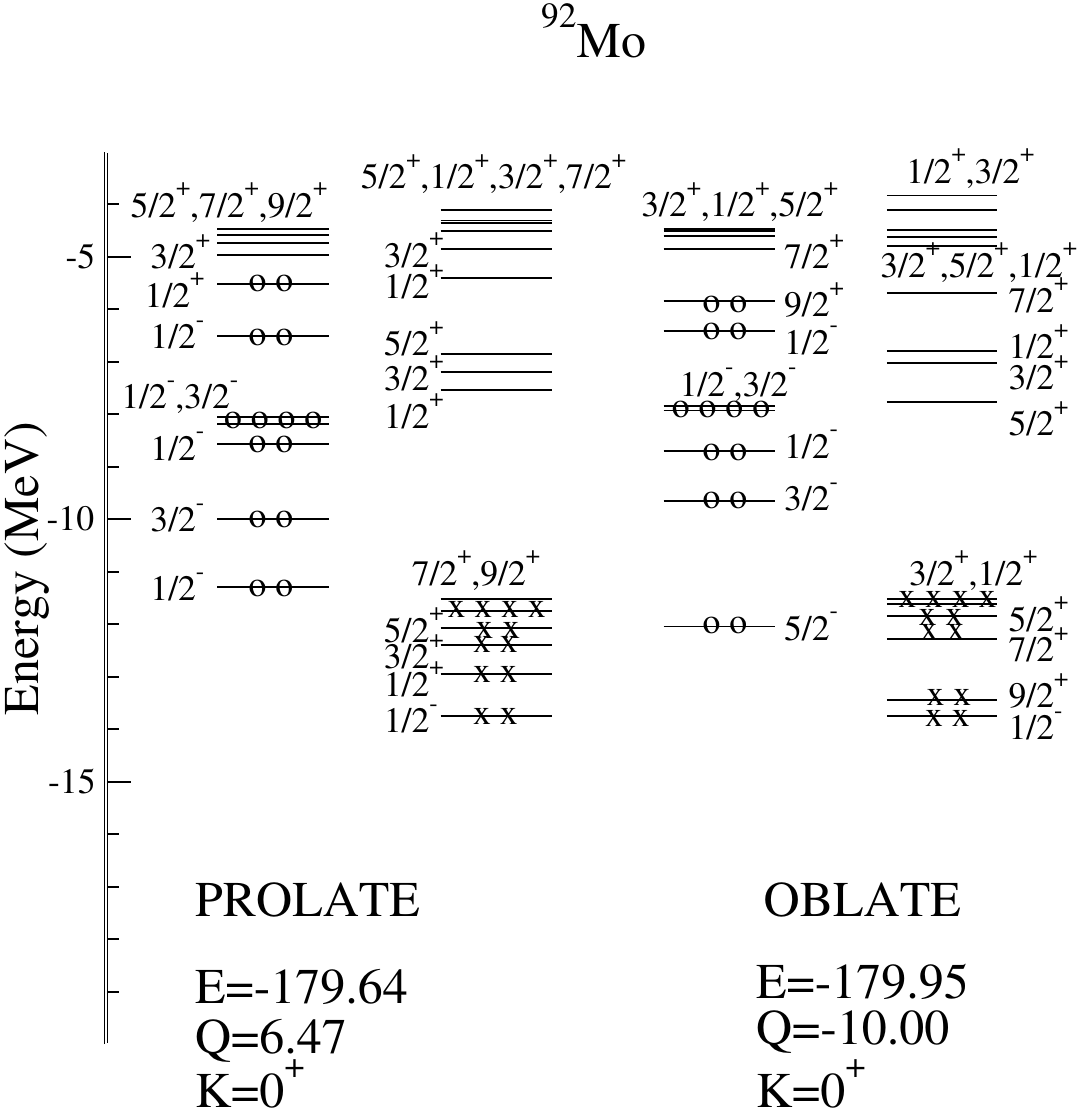}
       \caption{HF single-particle spectra for $^{92}$Mo  corresponding
        to lowest energy prolate and oblate configurations. 
        In the figure, circles represent
        protons and crosses represent neutrons. The HF energy E in MeV, mass 
        quadrupole moment Q in units of the square of the oscillator length
        parameter and the total azimuthal quantum
        number K are given in the figure.
}
\label{fig1}
\end{figure}

As described in Sect. II, we first obtain the lowest HF configuration by performing an axially 
symmetric HF calculation. Then, various
excited configurations are obtained by making particle-hole excitations
over this lowest HF configuration. The lowest energy prolate and oblate HF
energy spectrum is shown in Fig. \ref{fig1}. For both configurations, the
intrinsic quadrupole moments are small indicating that this nucleus has small
deformation. Both the prolate and oblate configurations are almost degenerate,
with the oblate solution appearing lower by around 0.3 MeV. Hence, there is good 
mixing between prolate and oblate configurations.

Similar prolate-oblate mixing has been seen in the four stable even-even
$^{94,96,98,100}$Mo isotopes studied in Refs. \cite{JPG2024,MDP1,MDP2} establishing 
the fact that the Mo isotopes are transitional nuclei. By particle-hole excitations 
over the lowest energy configurations, we have generated 16 intrinsic states of prolate
shape and 12 intrinsic states of oblate shape for $^{92}$Mo. Good angular momentum 
states are projected from each of these intrinsic states, and then a band mixing
calculation is performed as described before. 

\begin{figure}
\includegraphics[height=8.8 cm]{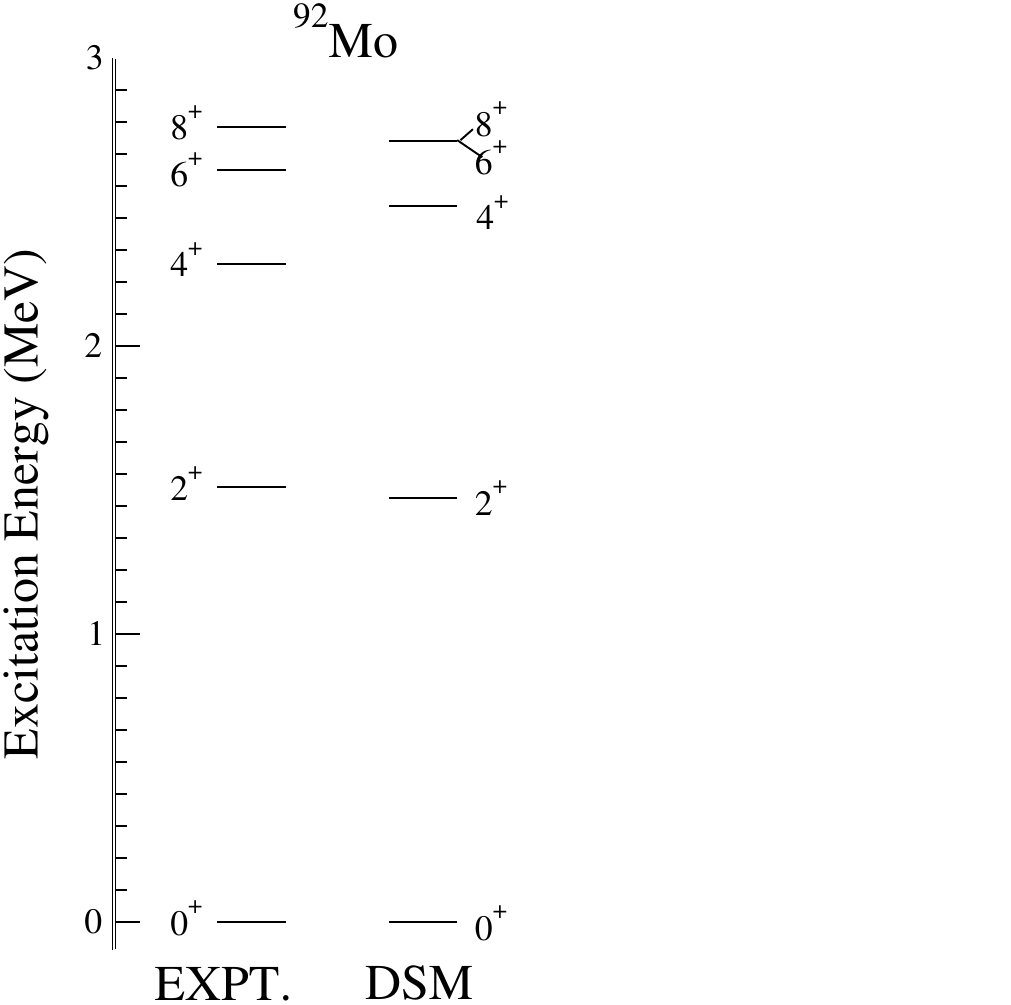}
        \caption{The ground band observed for $^{92}$Mo is compared with the DSM 
        predicted values. The experimental data are taken from \cite{nndc}
}
\label{fig2}
\end{figure}

The calculated yrast-levels are compared with experimental data in Fig. \ref{fig2}. 
As can be seen from this figure the agreement is reasonable.
Since $^{92}$Mo is a neutron closed-shell nucleus with $N=50$, the separation
of ground state $0^+$ to the first excited state $2^+$ is large, a feature that 
is nicely reproduced in our calculation (see Fig. \ref{fig2}). 
Then, there are three close-lying levels $4^+$, $6^+$, and $8^+$, which are also
reasonably well reproduced.

For further testing of the reliability of the wave functions generated by the DSM method, 
we have also calculated the magnetic moment of $2^+_1$ level as well as the $B(E2)$ values 
for some of the low-lying excited states. Using the bare gyromagnetic ratios for protons 
and neutrons, the calculated value for the magnetic moment of the $2^+_1$ state is
2.8 $\mu_N$ which, compared to the experimental value 2.3 $\mu_N$, shows that the agreement 
is reasonable. A slight adjustment of the values of gyromagnetic ratios would have yielded 
a much better agreement. Such adjustments have been made in many theoretical studies \cite{geff}. 
We note that, the magnetic moments for higher $J$-states have not been measured. 

With respect to the B(E2) values of $^{92}$Mo, two transitions, namely, the 
$2^+_1 \rightarrow 0^+_1$ and the $4^+_1 \rightarrow 2^+_1$, with values of 8.4 Wu and
$<$ 24 Wu, respectively, have been experimentally observed. The calculated values for 
these two transitions are 8.5 Wu and 17.9 Wu, in good agreement with experimental data. 
We mention that, in the calculation of B(E2) values, we have chosen the effective 
charges for protons and neutrons to be $e_p = 1.6e$ and $e_n = 1.0e$ as in our previous 
studies \cite{ks-book,JPG2024}.

Relying on the DSM wave functions tested as described above, we proceeded further and
calculated the coherent and incoherent neutrino- and antineutrino-$^{92}$Mo scattering 
differential cross sections as discussed below. Moreover, by employing the DSM wave
functions derived previously \cite{JPG2024,MDP2}, the incoherent antineutrino scattering 
cross sections for the even $^{94,96,98,100}$Mo isotopes were also computed.
We remind that the corresponding neutrino scattering cross sections on the latter 
even Mo isotopes have been published in Refs. \cite{JPG2024,MDP2}. 

\subsection{ Incoherent Cross Section Calculations for $^{92}$Mo $\nu$-Detector}

For the incoherent channel the calculations start from the double differential cross 
sections $d^2\sigma / d\Omega d\omega$ of Eq. (\ref{eq.11}) computed state-by-state 
within the DSM method. Subsequently, by integrating over the angles $\theta$ and $\phi$, 
the cross sections $d\sigma /d\omega$ are evaluated.

Because cross sections for high-J$^\pi$ multipole states (including also all the excited 
$0^+$ states) are found to be negligibly small, the incoherent differential cross sections 
discussed in this work, originate from the low-lying multipole states for specific incoming 
neutrino energies $\varepsilon_\nu$.

\begin{figure*}
\includegraphics[width=07.0cm]{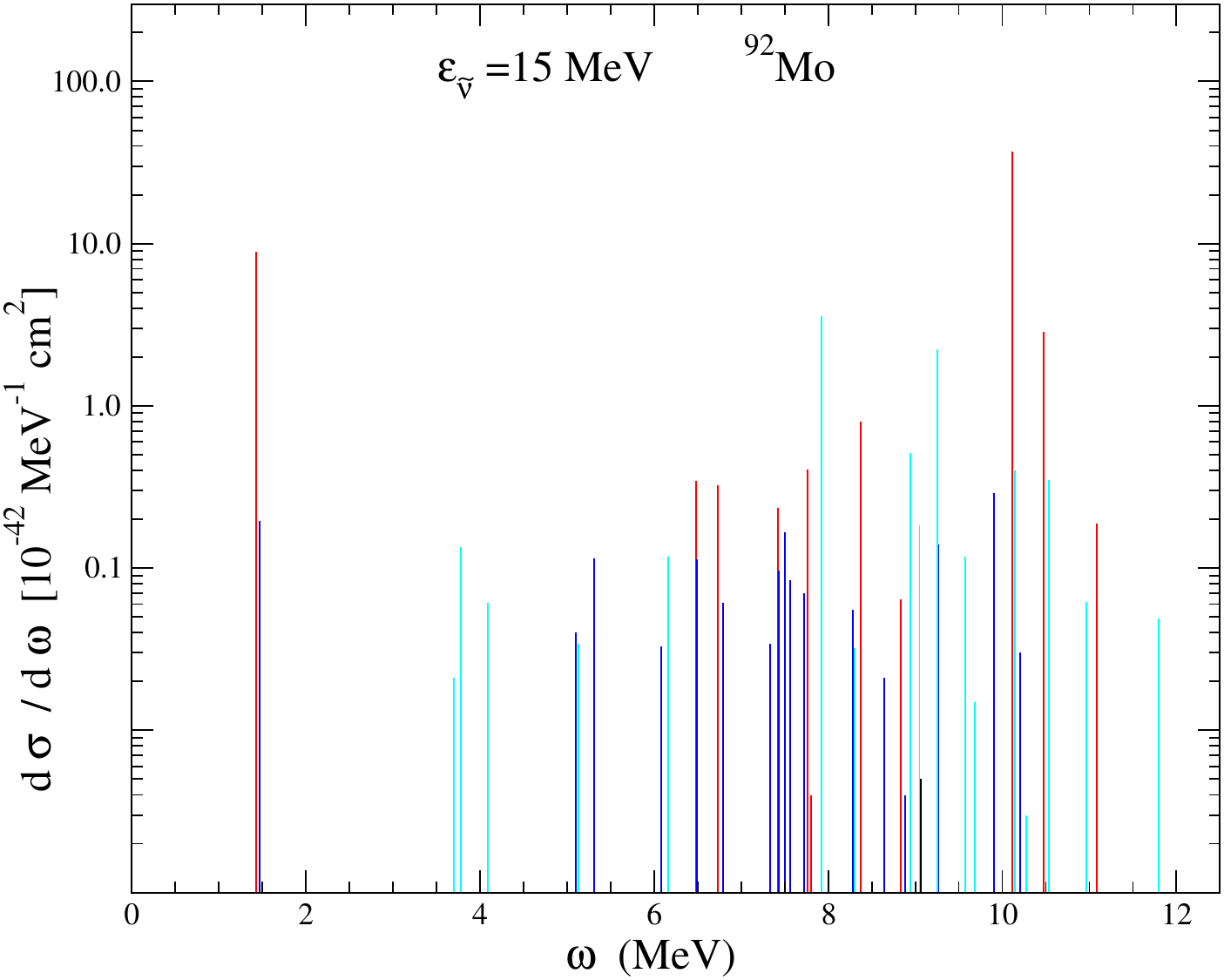} 
\hspace*{-0.05cm}
\includegraphics[width=07.0cm]{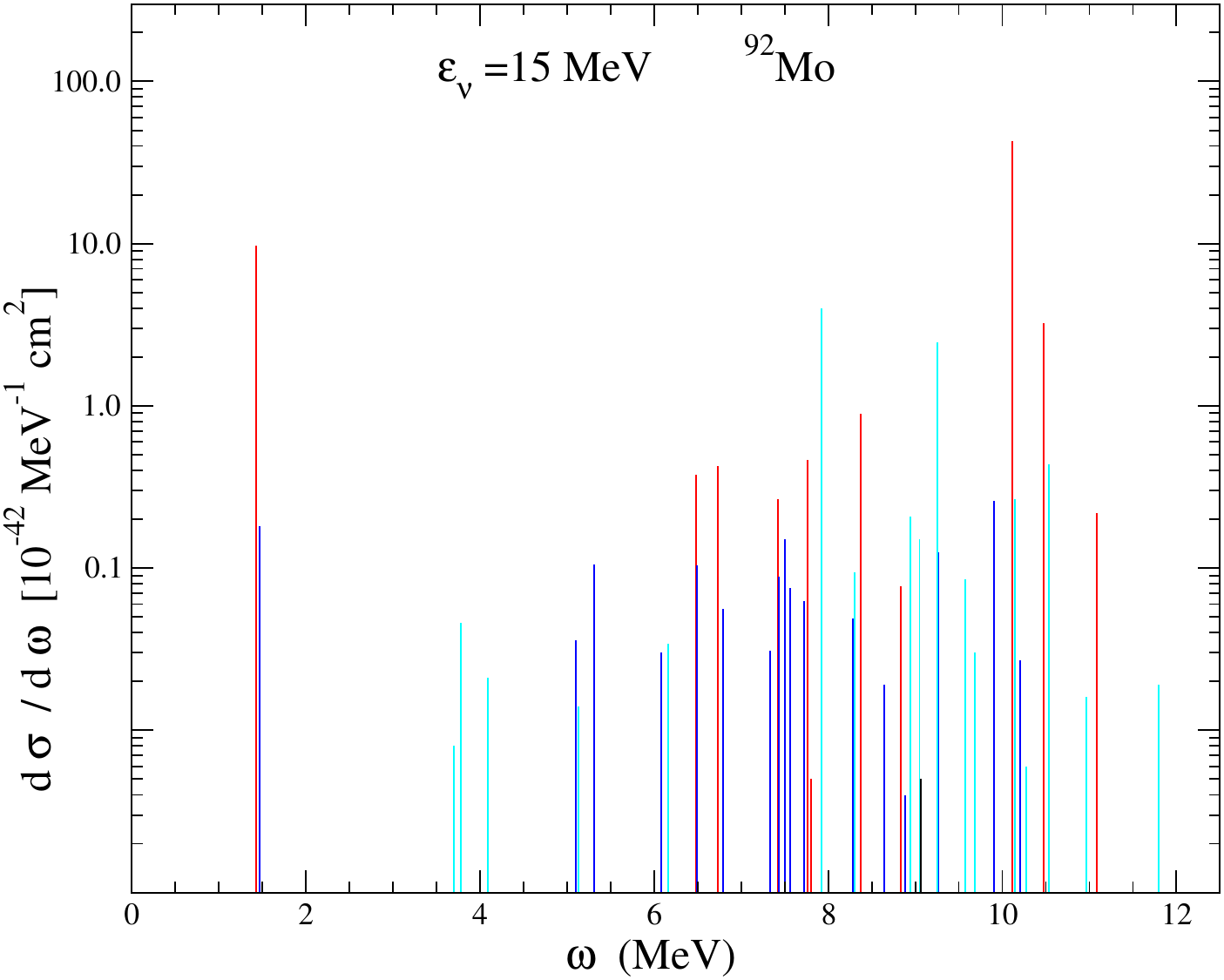} \\
\includegraphics[width=07.0cm]{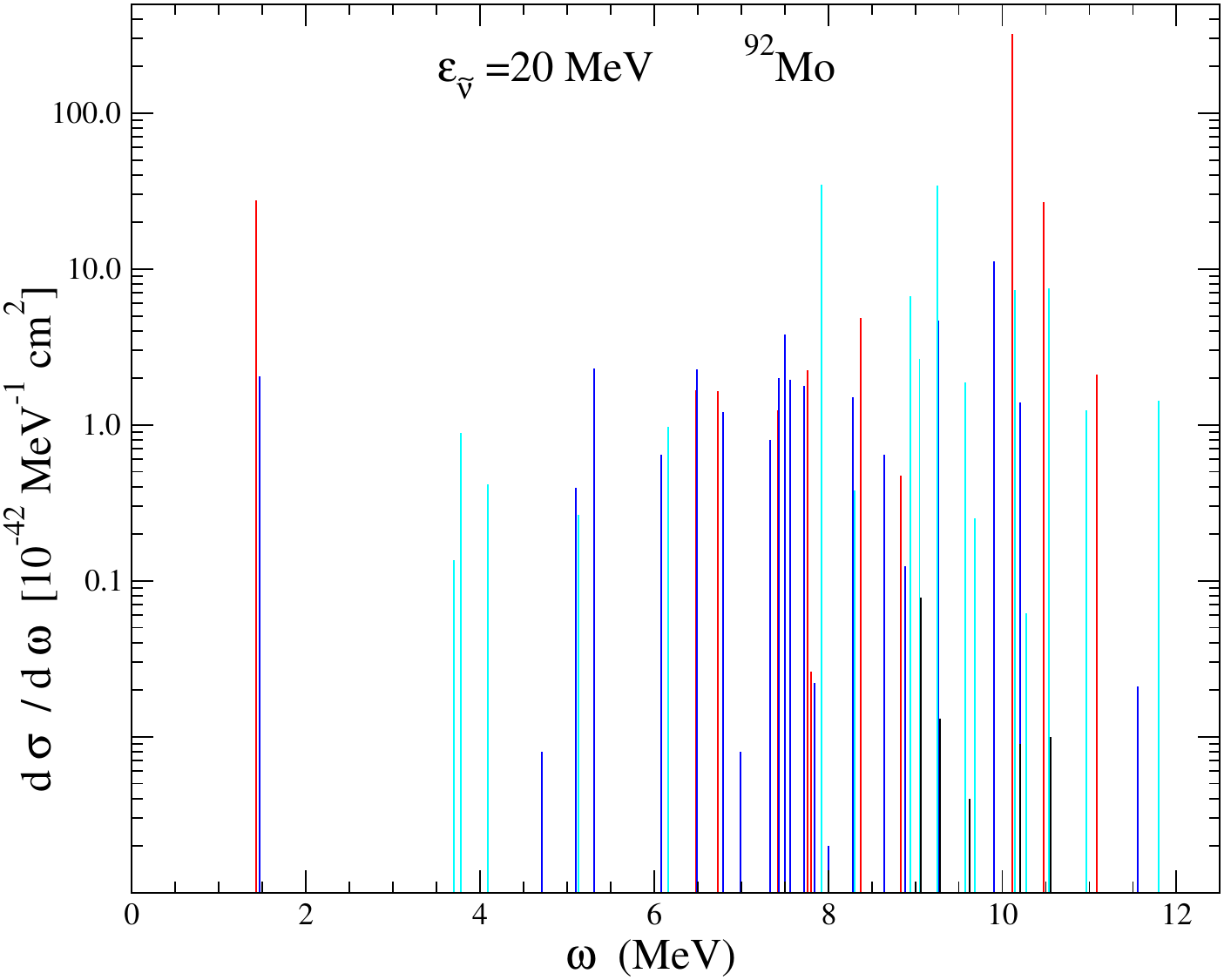} 
\hspace*{-0.05cm}
\includegraphics[width=07.0cm]{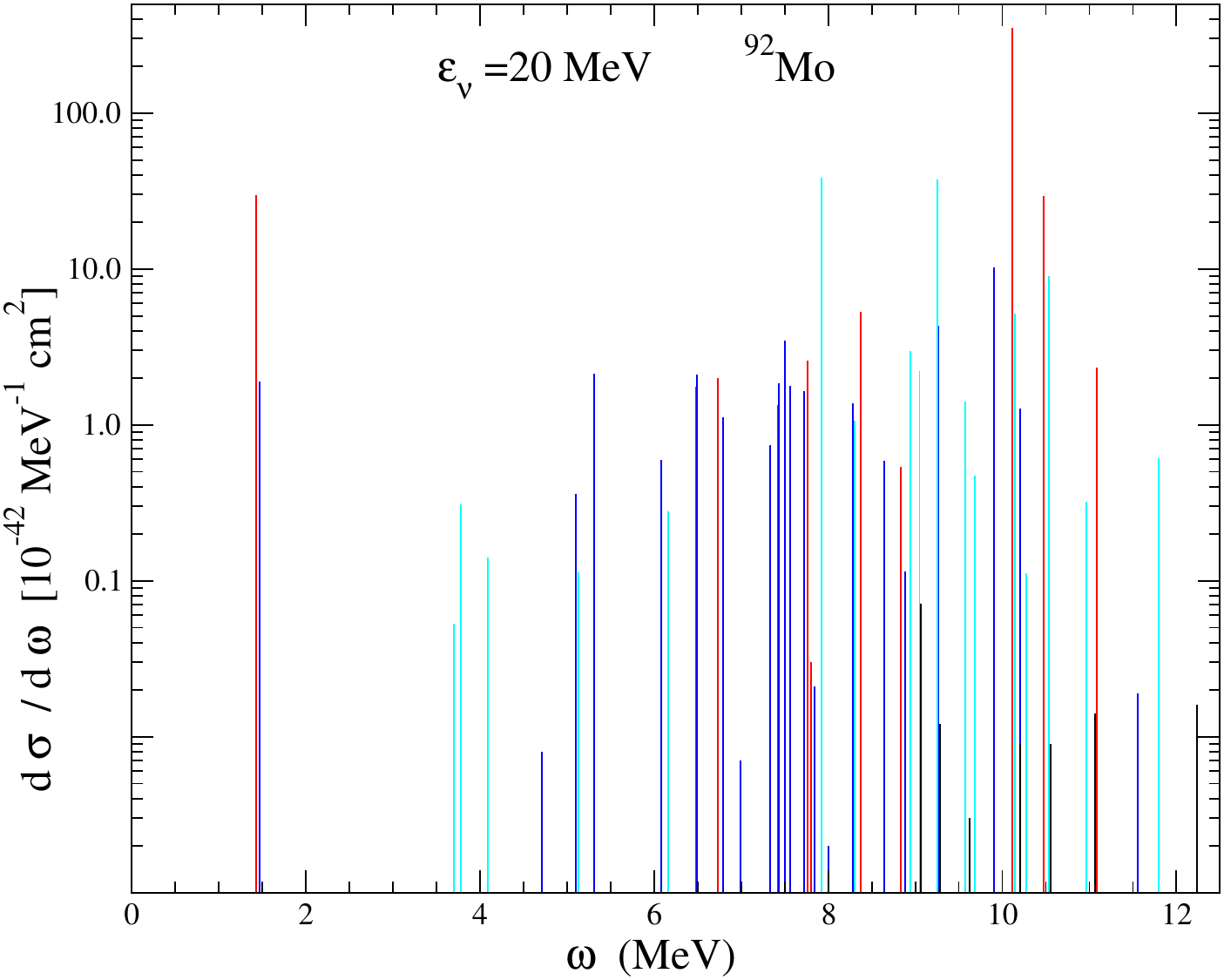}
\caption{ The differential cross section as a function of the excitation energy $\omega$
for $^{92}$Mo isotope:(a) Upper panel, at incoming antineutrino (left) energy 
$\varepsilon_{\tilde{\nu}} = 15$ MeV and neutrino (right) energy $\varepsilon_\nu = 15$ MeV, and 
(b) Lower panel, at incoming antineutrino (left) energy $\varepsilon_{\tilde{\nu}} = 20$ MeV and
neutrino (right) energy $\varepsilon_\nu = 20$ MeV, for different excited states of the
target nucleus. The contribution of the excitation to $J=1^+$ state is represented in red, to 
$J=2^+$ in blue, to $J=1^-$ in black, and to $J=2^-$ in cyan. The left side figures are for
antineutrino scattering and the right side ones for neutrino scattering. }
\label{fig3}
\end{figure*}

In Fig. \ref{fig3}, the results for the cross sections $d\sigma/d\omega$ are shown for
the incoming neutrino energy of $\varepsilon_\nu = 15$ MeV (or antineutrino energy
$\varepsilon_{\tilde{\nu}} = 15$ MeV). We have chosen 
to illustrate the dependence of $d\sigma/d \omega (\omega)$ for the value of incoming 
neutrino energy $\varepsilon_\nu = $ 15 MeV (or antineutrino $\varepsilon_{\tilde{\nu}} = 15$ MeV),
because it is slightly higher 
than the maximum energy of solar neutrinos ($\nu_e$) and reactor antineutrinos ($\tilde{\nu}_e$). 
We further mention that, in supernova (SN) neutrino simulations, the mean energies chosen 
for $\nu_e$ and $\tilde{\nu}_e$ (low-energy SN neutrinos) are between 12 MeV and 15 MeV. 
This means that, roughly speaking, our results may be used for estimations in solar, reactor,
and low-energy supernova neutrinos. 

In the present study of the $^{92}$Mo isotope, for $\varepsilon_\nu = 15$ MeV 
(or $\varepsilon_{\tilde{\nu}} = 15$ MeV), we found 
(see Fig. \ref{fig3}) that in both reactions, $^{92}Mo(\nu, \nu^\prime)^{92}Mo^\star$ 
(left panels) and $^{92}Mo(\tilde{\nu}, \tilde{\nu}^\prime)^{92}Mo^\star$ (right panels),
the pronounced peaks correspond to $J^\pi=1^+$, $2^+$, $1^-$ and $2^-$. Note that, as has also
been shown in Ref. \cite{Tsakstara-tsk}, the dominance of the incoherent peaks changes rather 
significantly with the chosen incoming neutrino/antineutrino energy. This can be implied
by comparing the upper panel of Fig. \ref{fig3}) (resulted with $\varepsilon_\nu = 15$ MeV
or $\varepsilon_{\tilde{\nu}} = 15$ MeV) and the lower panel (resulted with 
$\varepsilon_\nu = 20$ MeV or $\varepsilon_{\tilde{\nu}} = 20$ MeV).
As can be seen from Fig. \ref{fig3}, the incoherent cross section originates mostly 
from the $1^+$ states and, specifically, from the two $1^+$ states at $1.43$ MeV and 
at $10.47$ MeV. These two excited states generate almost the total cross section due to
the low-lying excitations of the $^{92}$Mo isotope. 

\begin{figure*}
\begin{center}
\includegraphics[width=06.0cm]{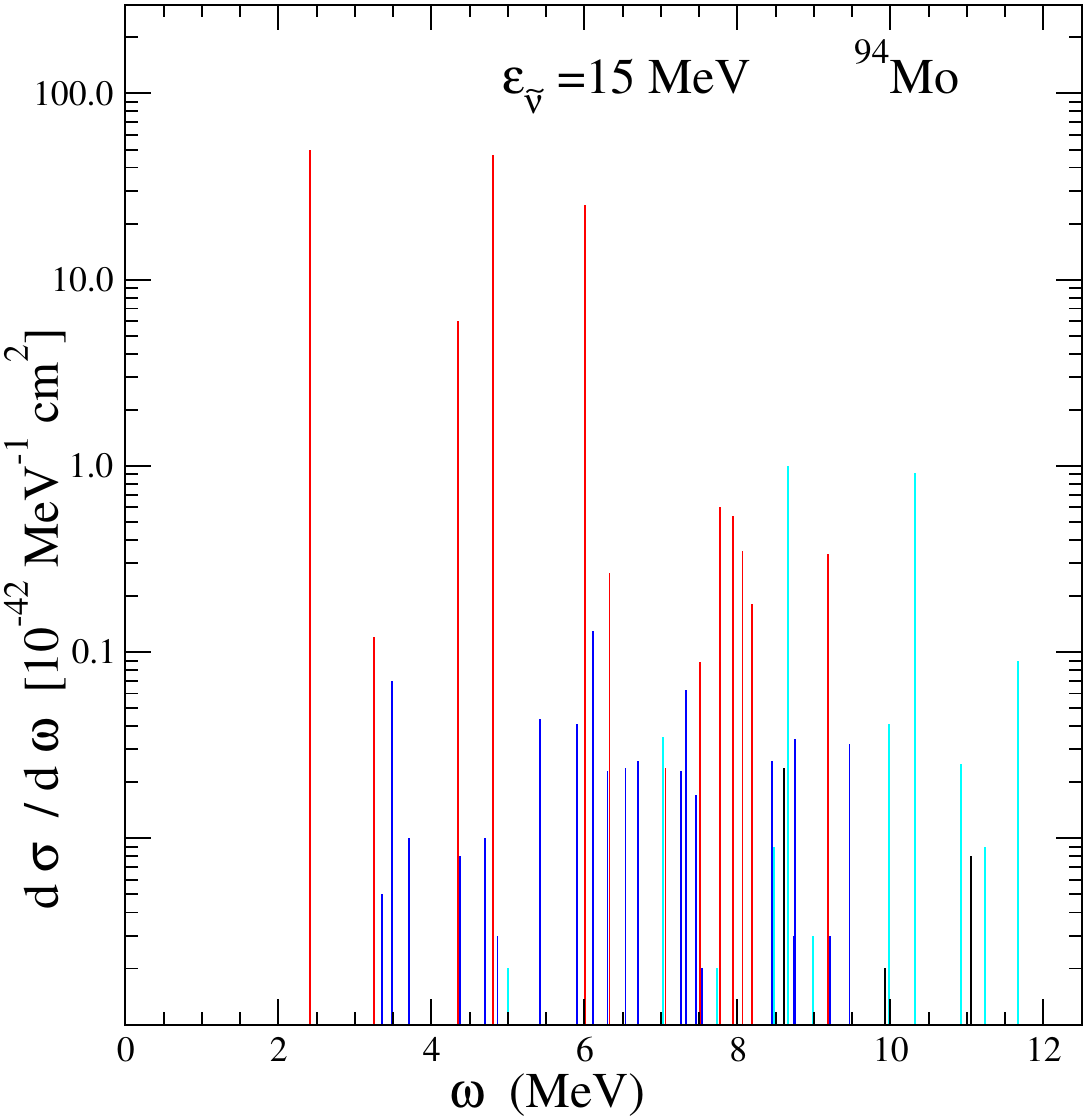}\hspace{0.2cm} 
\includegraphics[width=06.0cm]{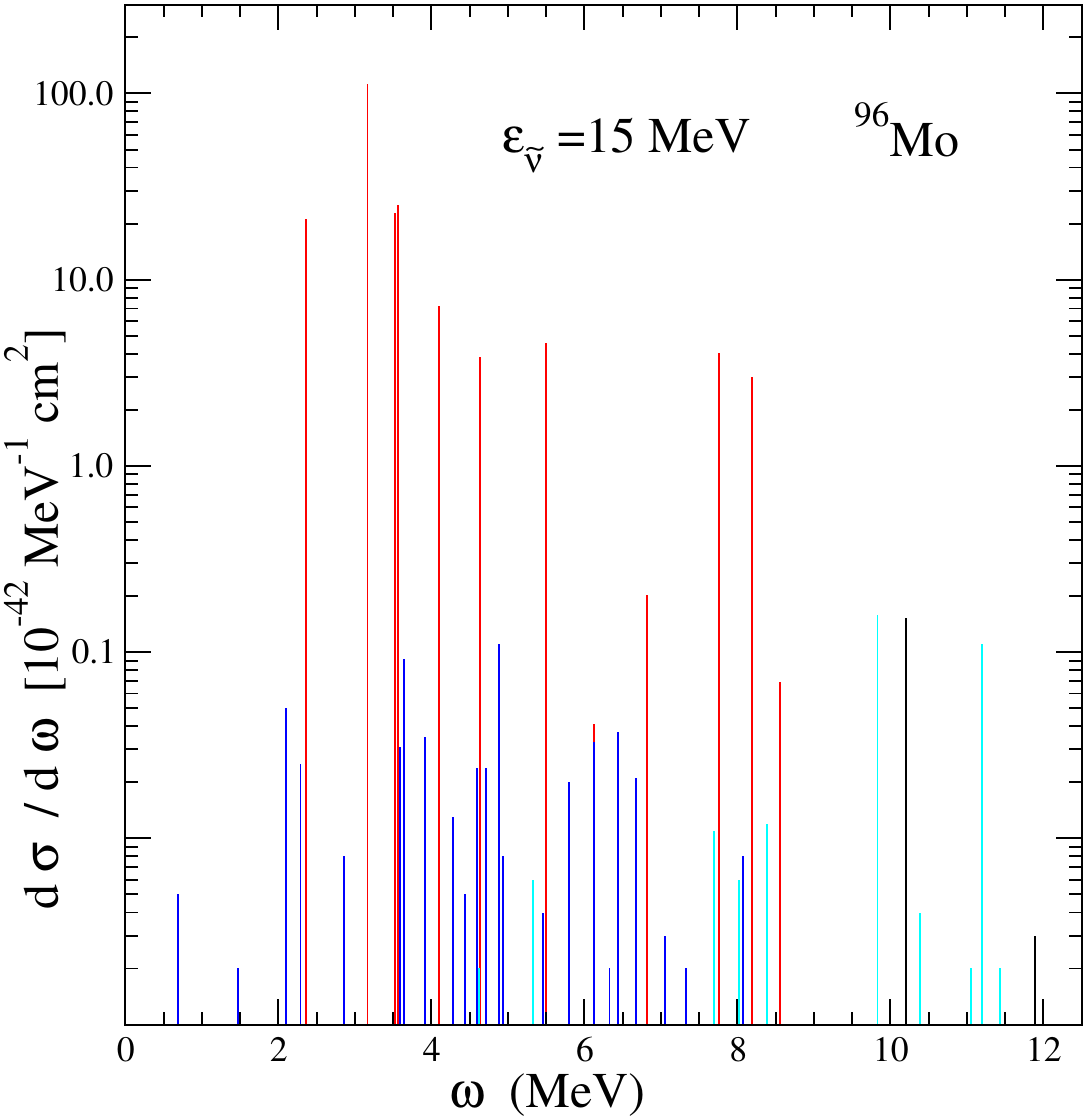} \\
\includegraphics[width=06.0cm]{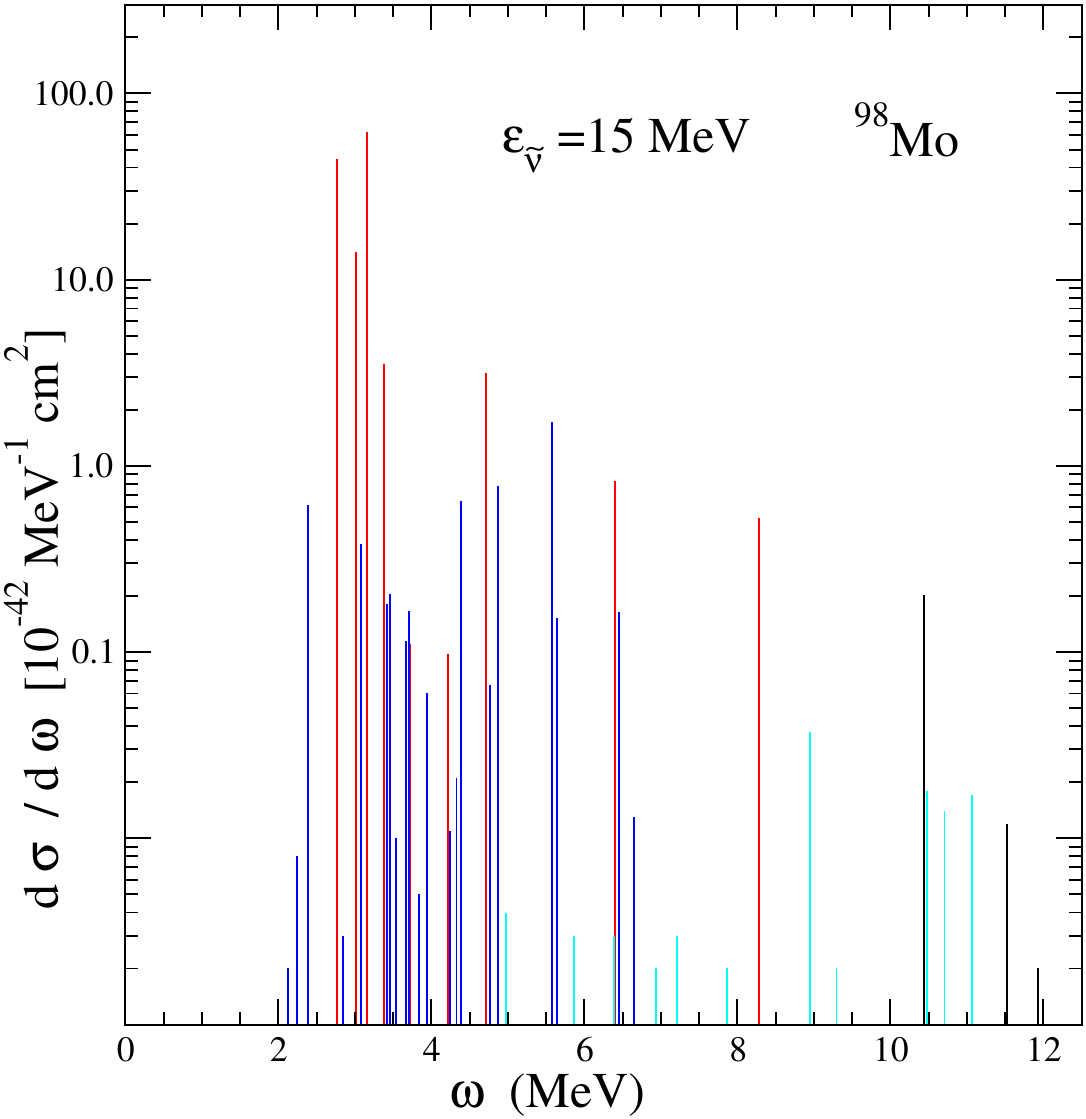} \hspace{0.2cm} 
\includegraphics[width=06.0cm]{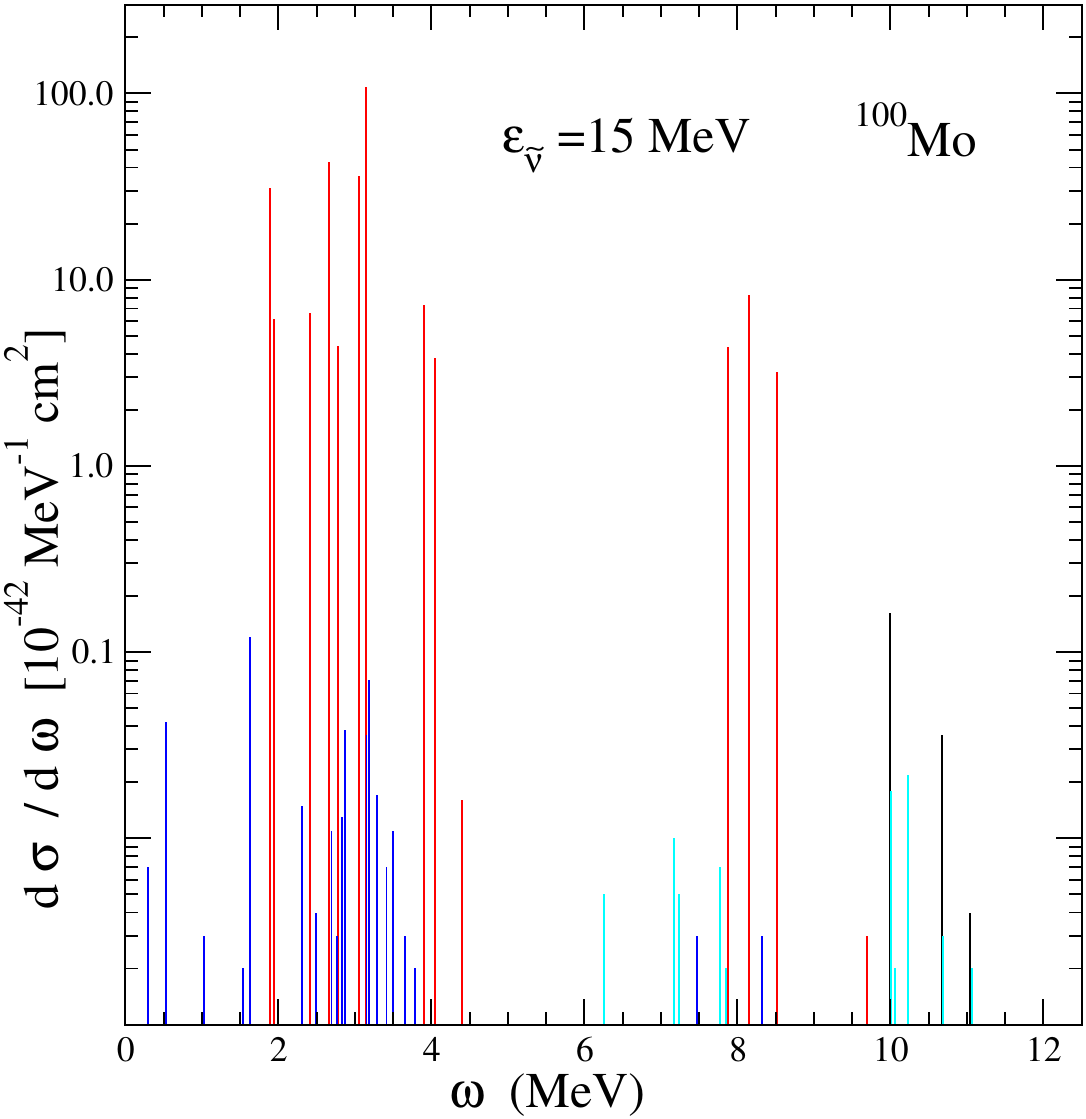}
\caption{The differential cross section as a function of the excitation energy $\omega$ for 
$^{94,96,98,100}$Mo at incoming antineutrino energy $\varepsilon_{\tilde{\nu}} = 15$ MeV
for different excited states.  The contribution of the excitation to $J=1^+$ states is 
represented in red, to $J=2^+$ in blue, to $J=1^-$ in black, and to $J=2^-$ in cyan.}
\label{fig4}  
\end{center}                
\end{figure*}

The results illustrated in Figs. \ref{fig3} and \ref{fig4} (also those published in Refs. 
\cite{JPG2024,MDP1,MDP2}) show that the Mo-isotopes present rich responses in the excitation 
energy range $\omega \le 15-20$ MeV. This energy region includes transitions to the bound 
nuclear states and is relevant for solar neutrinos, reactor neutrinos, and geo-neutrinos, 
but also for the low-energy supernova neutrinos. Thus, the calculated (anti)neutrino-Mo 
isotopes cross sections are suitable for use also in astrophysical neutrino simulations 
(e.g. interpretation of neutrino oscillations, neutrino properties, etc.).

It is worth mentioning that in these calculations we used the quenching factor 0.35 (defined by 
Kay et al. \cite{Kay}), which had also been utilized in our previous studies \cite{JPG2024,MDP2}).

By utilizing the neutrino energy distributions of the important neutrino sources mentioned above,
folded (differential) cross sections may be obtained that represent the response of the nuclear 
detector of interest to the specific neutrino sources (see e.g. \cite{Tsakstara-tsk,qrpa-1}). 

\section{Comparative analysis of incoherent cross sections}

The cross section results presented here for (anti)neutrino scattering on $^{92}$Mo, may now 
be discussed in conjunction with the results we have reported earlier in Refs. \cite{JPG2024,MDP2} 
for the four stable even-even Mo isotopes with A=94, 96, 98, and 100. As is known, the natural 
abundances of the stable even-even molybdenum isotopes are: for the $^{92}$Mo 15.86\%, for the
$^{94}$Mo 9.12\%, for the $^{96}$Mo 16.50\%, for the $^{98}$Mo 23.75\%, and for the $^{100}$Mo 
9.62\%, as mentioned in Section I. 
This means that, for a pure (non-enriched) molybdenum neutrino-detector material, the even-even 
isotopes represent nearly the 75\% of the detector mass. On the other hand, the only stable odd Mo 
isotopes are the $^{95}$Mo and $^{97}$Mo, with natural abundances of 15.70\% and 9.45\%, respectively.
  
Even in the incoherent part,
the difference is no more than 10\% as seen from the numbers in Table \ref{Table1} 
and also from Fig. \ref{fig3}. 

We will now consider the incoherent part in more detail, as this is significant in the context 
of BSM, and consider only the antineutrino scattering cross sections.
\begin{figure*}
\includegraphics[width=12.5cm]{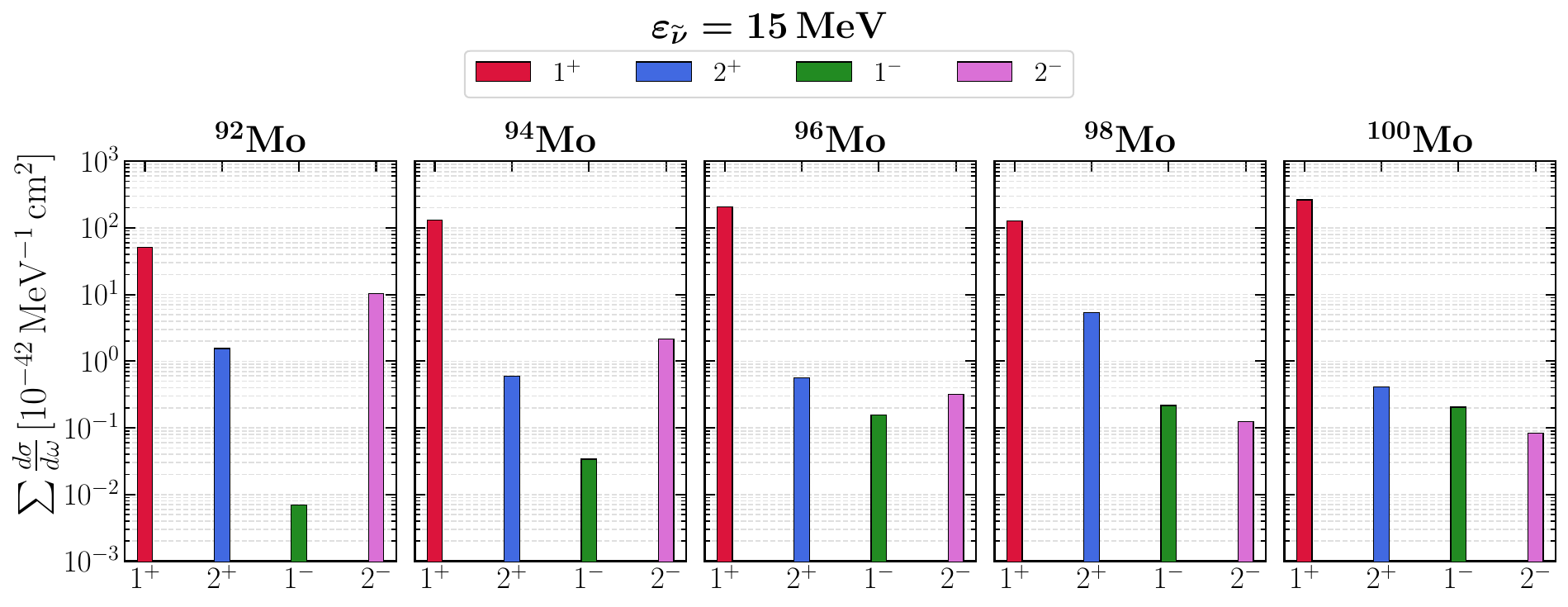} \\
\includegraphics[width=12.5cm]{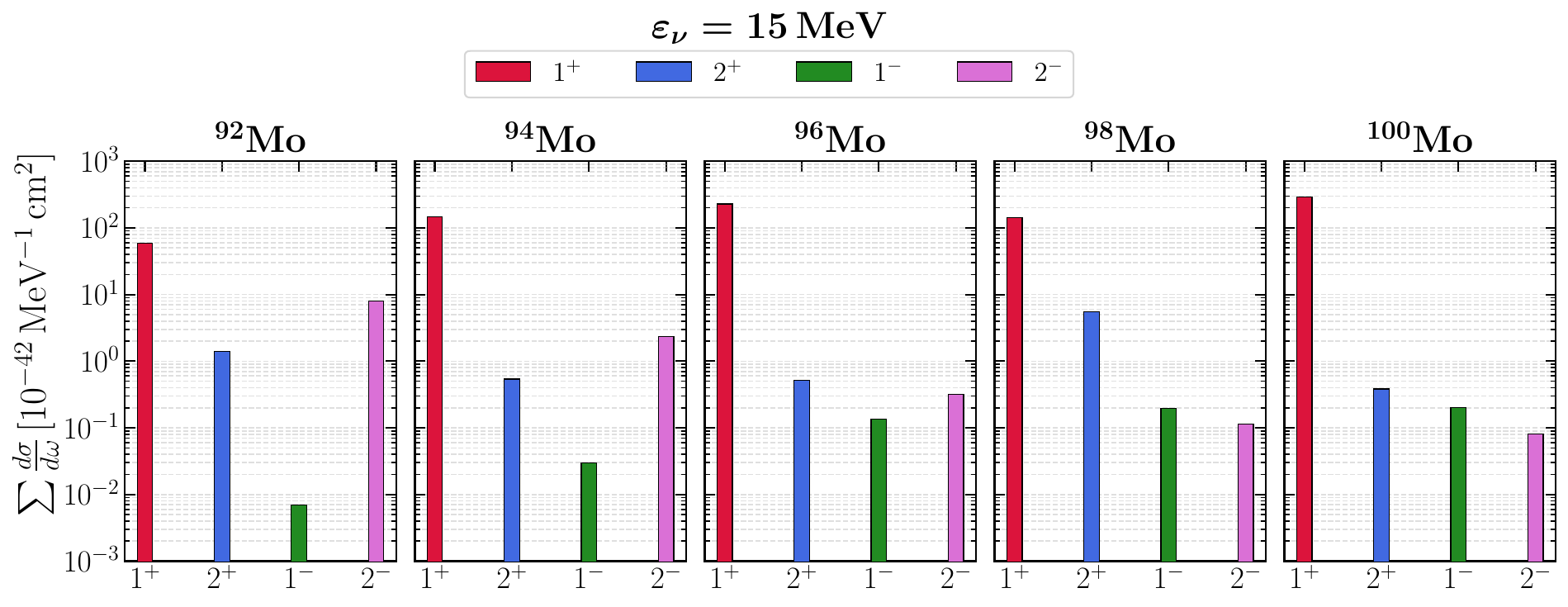}          
\caption{ Summed antineutrino-nucleus (upper panel) and neutrino-nucleus (lower panel) differential 
scattering cross sections for the $1^+$, $2^+$, $1^-$ and $2^-$ states for the five isotopes 
$^{92,94,96,98,100}$Mo. }
\label{fig5}
\end{figure*}
%
%
\begin{figure*}
\includegraphics[height=8.0cm]{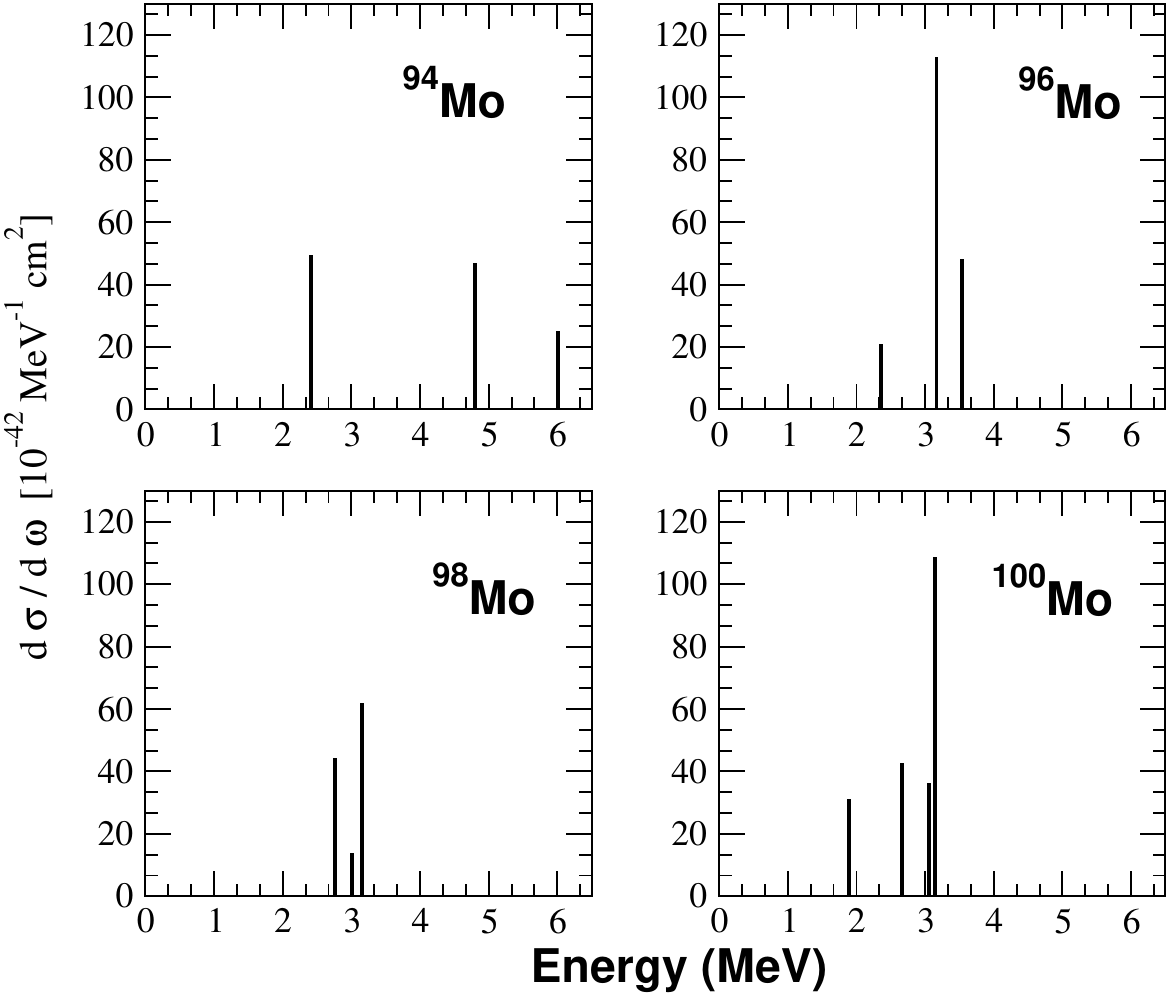}
\caption{The antineutrino nucleus differential scattering cross section for the dominant $1^+$ 
states as a function of their excitation energy for the four isotopes $^{94,96,98,100}$Mo}.
\label{fig6}
\end{figure*}

The results illustrated in Fig. \ref{fig3} for antineutrino scattering on $^{92}$Mo 
are similar to those for $^{94,96,98,100}$Mo shown in Fig. \ref{fig4}.

For the sake of comparison, in Fig. \ref{fig5} we illustrate the scattering differential 
cross sections leading to the individual multipolarities $1^+$, $2^+$, $1^-$, and $2^-$. 
It is clear from Fig. \ref{fig5} that most ($ >90$\%) of the cross section originates
from the $1^+$ excitations. This implies that, the neutrino scattering on the natural Mo 
detector medium is expected to probe mostly the Gamow-Teller like $1^+$ states. 

Proceeding further, in Fig. \ref{fig6}, we show the cross section for the dominant $1^+$ 
states as a function of their excitation energies. By combining the results in Fig. \ref{fig6} 
and Fig. \ref{fig3}, we conclude that, the cross section strength is mainly concentrated 
in three to four pronounced $1^+$ levels and that this holds for all the five stable even-even 
Mo isotopes. 

We note that, the cross sections for antineutrinos in the above figures differ by 10\% (see 
Table \ref{Table1}) from the corresponding neutrino-Mo isotopes cross sections of Ref. \cite{JPG2024,MDP2}.

\section{Coherent, Total Cross Sections and the Ratio $\rho$}

\begin{table*}
\caption{\label{Table1} Coherent, incoherent and total differential cross sections (in units 
of $10^{-42}$ MeV$^{-1}$ cm $^2$) of neutrino and antineutrino scattering off the $^{92,94,96,98,100}$Mo 
detector isotopes. The portion of the coherent into the total cross section is also listed. 
}
\begin{center}
\begin{tabular}{l l r c c c c r}
\hline
\hline \\ [0.005ex]
Isotope{\hspace*{1.0 cm}} &  Coherent{\hspace*{0.5 cm}}  & \multicolumn{2}{c}{
Incoherent{\hspace*{1.0 cm}} } & 
\multicolumn{2}{c}{Total Cross Section{\hspace*{1.0 cm}}} & \multicolumn{2}{c}{Ratio $\rho $  (\%) } \\
        &           & \multicolumn{1}{c}{($\nu$)} & \multicolumn{1}{c}{($\tilde{\nu}$)}  & 
        ($\nu$) & {\hspace*{0.1 cm}}($\tilde{\nu}$)        &  
        {\hspace*{0.1 cm}}($\nu$) & {\hspace*{0.5 cm}}($\tilde{\nu}$)  \\
\hline \\ [0.01ex]
$^{92}$Mo  &   998.6  &  67.7  &  63.1 & {\hspace*{0.5 cm}}1066.3 &  1061.7 & 93.65  & 94.05 \\
$^{94}$Mo  &  1165.0  & 150.4 & 132.7  & {\hspace*{0.5 cm}}1315.4 &  1297.7 & 88.57  & 89.77 \\
$^{96}$Mo  &  1338.5  & 230.0 & 206.1  & {\hspace*{0.5 cm}}1568.5 &  1544.6 & 85.34  & 86.66 \\
$^{98}$Mo  &  1506.2  & 147.7 & 134.2  & {\hspace*{0.5 cm}}1653.9 &  1640.4 & 91.07  & 91.82 \\
$^{100}$Mo &  1692.9  & 290.0 & 263.5 & {\hspace*{0.5 cm}}1982.9 &  1956.4 & 85.37  & 86.53 \\
\hline
\hline
\end{tabular}
\end{center}
\end{table*}

In Table \ref{Table1}, we list the coherent cross section, the sum of the cross section 
contributions coming from nearly all the excited states, called incoherent cross sections,
and the total (= coherent + incoherent) cross section for both cases: neutrino-Mo and
$\tilde{\nu}$-Mo isotopes neutral current scattering reactions. By exploiting the results of 
Table \ref{Table1} we compare the coherent, incoherent, and total cross sections for the 
five stable even-even Mo-isotopes and calculate the ratio of the coherent divided by the 
total cross section, i.e., the quantity
\begin{equation}
\rho = \frac{Coherent}{Total} \, .
\end{equation}
This is interesting from an experimental point of view since, in general, the quantity $\rho$
represents the portion of the total cross section that is measured in the CE$\nu$NS experiments.

It is important to note that the coherent cross sections for neutrino-nucleus and antineutrino-nucleus  
scattering are identical (only in the incoherent cross sections are different). Furthermore, as
can be readily found the coherent cross sections show a good correlation to the square of the neutron 
number ($\sim N^2$) of each isotope. 

It should be stressed that the results listed in Table \ref{Table1} for Mo-(anti)neutrino scattering show 
Standard Model predictions. As has been observed by the CE$\nu$NS experiments, the number of the 
measured events is larger (both for the neutrino and antineutrino sources) than the SM predictions.
Thus, by invoking BSM physics one may choose several scenarios, such as neutrino magnetic moments, 
neutrino millicharges, light neutrino bosons, etc. Then, in view of future measurements of CEvNS 
events on Mo detectors one may compare our SM predictions with the SM+BSM ones on the one side and 
the experimental data on the other. 

In the case of the Ge detector medium, such calculations have been
published and the reader is referred to Refs. 
\cite{Papoulias-tsk-2018,solarnu,CONUS-PRL-2023,Giunti-PLB-2025}. 
It is worth mentioning that, in general, even though the neutral-current neutrino-nucleus 
scattering cross sections evaluated here, are substantially smaller than those of the 
charged-current neutrino-nucleus scattering ones, the former may provide important information 
that is attributed to the following reasons. In the low-energy range of our calculations,
$\varepsilon_\nu \lesssim 15$ MeV (also for energies $\varepsilon_\nu \lesssim 100 $MeV), 
neutrinos are not able to produce the massive leptons ($\mu$, $\tau$) in the detector, which 
means that in low-energy neutrino detectors the portion of neutrinos that will participate 
in charged-current scattering is limited. 

Moreover, charged-current antineutrino scattering, in medium heavy nuclei (as is the case of 
Mo-isotopes) as well as heavy nuclei, is suppressed due to Pauli blocking effects. Hence, only 
electron-neutrino charged-current reactions are significant for such detectors. On the other
hand, in neutral-current scattering of (anti)neutrinos on, e.g. Mo detectors studied in this 
work, specifically the heavy flavor neutrinos ($\nu_x$, with $x=\mu, \tau$) can also be detected. 

Regarding the main uncertainties in neutral current neutrino-nucleus scattering cross sections 
or event rates, in the case of the measured CEvNS channel (where the cross sections are essentially 
proportional to $N^{2}$) they generally stem from the weak neutron form factor, the neutrino flux, 
and the effects related to the specific detector employed. For intense neutrino fluxes required for 
CEvNS experiments, the number of neutrinos produced by the source is a major uncertainty and is 
expected to decrease as detectors like those used at the Spallation Neutron Source (SNS) at Oak 
Ridge \cite{COHERENT-Science,COHERENT-PRL} 
improve their measurement capabilities (future experiments aim for around a 2-3\% 
improvement). The source of uncertainty related to the effects of the specific detector employed
is that, the relationship between the detected nuclear recoil energy and the true recoil energy, 
is not perfectly known. On the other hand, uncertainties in the detector's quenching factor for 
nuclear recoils require improved measurements of these factors, which are crucial for achieving 
high precision. Currently, detector-related uncertainties are reduced through careful calibration, 
which is essential for precise measurements \cite{bsm-CMS,cenu-sm}. 

Focusing on the theoretical cross-section predictions, the main uncertainties for the coherent 
calculations originate from the nuclear model (random phase approximation, shell model, 
relativistic Fermi gas model, etc.) utilized to compute the weak nuclear form factor for the neutron 
(mainly) and proton distributions within the nucleus, which is a significant source of error. 
The calculations relying on various nuclear models usually lead to somewhat different predictions, 
especially when they are based on effective interactions or adjustments in single-particle parameters. 
Predictions based on the Shell Model or the deformed shell model employed in the present work may be 
uncertain due to factors like the uncertainties in the input values for magnetic transitions and the 
accuracy of the predicted shell-model states themselves. Advanced nuclear structure calculations 
usually employ shell models to better constrain the form factor uncertainties.

It is worth mentioning that the axial-vector coupling parameter $g_{A}$ used in nuclear physics is 
considered uncertain largely due to effects from the nuclear medium (in the complex nuclear environment,
$g_{A}$ is an effective coupling parameter) that quench the value from its free-nucleon value 
$g_{A}$ = 1.27641, which is highly precise. In many neutrino interaction models, the axial-vector 
coupling parameter is not precisely known and influences the cross-section calculations \cite{geff}.
Nuclear 
many-body effects and interactions with the nuclear medium lead to a quenching of the axial-vector 
coupling; i.e., the effective value is smaller than the free-nucleon value. We also mention that at 
higher neutrino energies (e.g., high-energy supernova neutrinos), the uncertainties of theoretical 
calculations can be more complex due to partial coherence. Finally, when including BSM physics, the 
challenge of distinguishing a signal of physics beyond the Standard Model from Standard Model 
theoretical uncertainties is one of the main hurdles for precision experiments; thus, the effort 
focuses on improving experimental precision while simultaneously reducing theoretical uncertainties.

\section{Summary and Conclusions}

In this work, relying initially on nuclear structure calculations performed within the deformed
shell model (DSM), we computed coherent and incoherent (anti)neutrino-$^{92}$Mo scattering
cross sections. For natural (non-enriched) detection media used in the CEvNS experiments,is
however, one 
should compute the CEvNS cross sections (events) coming from each of the stable Mo isotopes. 
Then, through the abundances of the stable isotopes in the natural molybdenum, one may obtain
the (theoretical) number of events separately for the coherent channel (to be compared 
with those measured by the Mo detector) and the incoherent channel. Theoretically, as shown 
in Table \ref{Table1}, one may estimate the portion of the neutral-current (anti)neutrino-nucleus scattering
events (expected to be measured by CEvNS experiments). In this work, we found that this ratio
is in between 86\% and 94\%.  

From an experimental viewpoint, the measured CEvNS events with accelerator pion decay-at-rest 
(DAR) neutrino beam on a Ge detector (at the SNS source) have been comparatively analyzed in 
conjunction with those measured by using a reactor anti-neutrino source on a common Ge detector
recently. These complementary 
results offer the possibility of a combined analysis of CEvNS results on a germanium target 
and a determination of important nuclear physics parameters \cite{Giunti-PLB-2025}.

The present cross sections of Mo-(anti)neutrino scattering are Standard Model predictions. 
In case of the existence of room for physics beyond the Standard Model (BSM), as shown from 
other detector media, one may choose various BSM theories (neutrino magnetic 
moments, the light neutrino bosons, the neutrino millicharges, etc.), for the interpretation 
of the CE$\nu$NS data \cite{{nth-4}}. 
Then, whenever measurements of CEvNS events on Mo detectors will come,
one may compare our SM predictions with the SM + BSM ones with the experimental 
data. Furthermore, these results may be discussed in conjunction with the 
published Ge detector results \cite{Papoulias-tsk-2018,solarnu,CONUS-PRL-2023,Giunti-PLB-2025}.

In conclusion, the potential of CEvNS may be further exploited by the next generation detectors 
with improved sensitivity, broader physics goals, and varied target materials. For example, 
the molybdenum detection medium studied in this work. Thus, future measurements of CEvNS 
events may significantly improve the precision of the relevant observables such as the neutron 
root-mean-square (rms) radius of the target Mo-isotopes, and other fundamental Standard Model 
quantities like the neutrino charge radius and the weak mixing angle originating from the 
Mo detector.

\section*{Acknowledgments}

One of us (TSK) wishes to thank Dr. Dimitrios Papoulias for fruitful discussions and
for relevant technical assistance during the preparation of the first manuscript.     

\renewcommand{\theequation}{A-\arabic{equation}}
\setcounter{equation}{0}
    
\section*{APPENDIX A}

The definitions of the eight multipole operators $\hat{M}_J$, $\hat{L}_J$,
$\hat{T}^{el}_J$, $\hat{T}^{mag}_J$, $\hat{M}^5_J$, $\hat{L}^5_J$,
$\hat{T}^{el5}_J$ and $\hat{T}^{mag5}_J$, where the subscript 5 refers to
the axial vector components of the hadronic current, are as follows:
\begin{equation}
        \begin{array}{l}
        \hat{M}^{Coul}_{JM}(q{\mathbf r}) = F^Z_1\hat{M}^J_M(q{\mathbf r}), \qquad 
        \hat{L}_{JM}(q{\mathbf r}) =\displaystyle{\frac{q_0}{q}}\hat{M}^{Coul}_{JM}(q{\mathbf r}),\\
\hat{T}^{el}_{JM}(q{\mathbf r})=\displaystyle{\frac{q}{M}}\left[F^Z_1\Delta^{\prime J}_M (q{\mathbf r})
+\frac{1}{2}(F^Z_1+2MF^Z_2)\Sigma^J_M(q{\mathbf r})\right],\\
i\hat{T}^{mag}_{JM}(q{\mathbf r})=\displaystyle{\frac{q}{M}}\left[F^Z_1\Delta^{ J}_M (q{\mathbf r})-
\frac{1}{2}(F^Z_1+2MF^Z_2)\Sigma^{\prime J}_M(q{\mathbf r})\right],\\
i\hat{M}^{5}_{JM}(q{\mathbf r})=\displaystyle{\frac{q}{M}}\left[F^Z_A\Omega^{J}_M (q{\mathbf r})+
\frac{1}{2}F^Z_A\Sigma^{\prime\prime J}_M(q{\mathbf r})\right],\\
-i\hat{L}^{5}_{JM}(q{\mathbf r})= F^Z_A \Sigma^{\prime\prime J}_M(q{\mathbf r}) ,\qquad
-i\hat{T}^{el5}_{JM}(q{\mathbf r})= F^Z_A \Sigma^{\prime J}_M(q{\mathbf r}),\qquad\\
\hat{T}^{mag5}_{JM}(q{\mathbf r})= F^Z_A \Sigma^{J}_M(q{\mathbf r}).
        \end{array}
        \label{eq.18}
\end{equation}
where we neglect the pseudoscalar form factor. In Eq. \ref{eq.18}, the first three and last 
multipole operators have normal parity, $\pi=(-)^J$ while others have abnormal parity, $\pi=(-)^{J+1}$.

\end{document}